\documentclass[twocolumn,nofootinbib,amsmath,amssymb,a4paper]{revtex4}

\usepackage{graphicx}% Include figure files
\usepackage{dcolumn}% Align table columns on decimal point
\usepackage{bm}

\input babarsym

\def\to           {\ensuremath{\rightarrow}}
\def\B            {\ensuremath{B}}
\def\Bz           {\ensuremath{B^{0}}}
\def\Bzb          {\ensuremath{\overline{B}^{0}}}
\def\piz          {\ensuremath{\pi^{0}}}

\def\CP      {\ensuremath{ CP }}

\def\C       {\ensuremath{ C }}
\def\S       {\ensuremath{ S }}
\def\acp     {\ensuremath{ {\cal A}_{CP} }}

\def\mes     {\ensuremath{M_{ES}}}
\def\deltat  {\ensuremath{\Delta t}}

\def\deltae  {\ensuremath{\Delta E}}
\def\deltamd {\ensuremath{\Delta m_d}}

\def\iab       {\ensuremath{ { ab^{-1} }}}

\def\epem      {\ensuremath{ { e^+ e^- } } }
\def\qqbar     {\ensuremath{ { q \overline{q} } } }

\def\iab{\ensuremath{\mathrm{ab^{-1}}}}

\def\BBbar{{\ensuremath{B\overline{B}}}}

\newcommand{\e}    [1]   {{\ensuremath{ \times 10^{{#1}}}}}
\newcommand{\su}   [1]   {{\rm{SU({#1})}}}

\begin{document}

\title{Measurement of $\alpha / \phi_2$ from $B\to \pi\pi$ decays}

\author{A. J. Bevan}
\email{a.j.bevan@qmul.ac.uk}
\thanks{This work is supported by PPARC and the DOE under contract DE-AC02-76SF00515.}
\affiliation{Physics Department, Queen Mary, University of London, E1 4NS, UK.}

\begin{abstract}
The current results on $\B\to \pi\pi$ decays and \su{2} constraints on the Unitarity Triangle angle $\alpha$ or $\phi_2$
from the \B-factories are summarised.  Based on these measurements, predictions of the isospin analysis constraints
at the end of the lifetime of both \B-factories are given.
\vspace{1pc}
\end{abstract}

\maketitle

\section{Introduction}

These proceedings are a summary of the experimental constraints on the
Unitarity Triangle angle $\alpha$ (or $\phi_2$)obtained from \B-meson decays
to $\pi\pi$ final states.
Measurements of $\alpha$ have recently been performed at the \B-factories, the \babar\
experiment~\cite{babar_nim} at SLAC and the Belle
experiment~\cite{belle_nim} at KEK using $\pi\pi$, $\rho\pi$ and $\rho\rho$ decays
which proceed via $b\to u$ transitions as discussed in Ref.~\cite{bevan2006}.
Details of the constraints on $\alpha$ from
$\rho\pi$ and $\rho\rho$ decays can be found in
Refs.~\cite{somovCKM,cavotoCKM,babarrhoprhomprlr12,babarrhoprhomr14,babarrhoprhomr15prelim,babar_btorhopi,belle_btorhopi,bellerhoprhom}.

\begin{figure}[!h]
\begin{center}
  \resizebox{8.5cm}{!}{\includegraphics{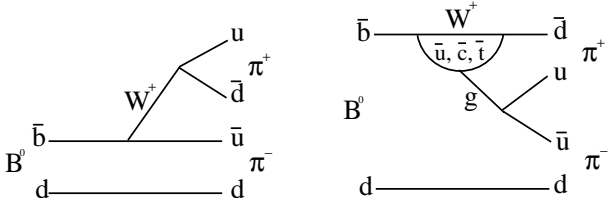}}
\end{center}
 \caption{Tree (left) and gluonic penguin loop (right) contributions to $\B\to\pi\pi$ decays.}
\label{fig:feynman}
\end{figure}

The decays $\B \to \pi\pi$ proceed mainly through a $\overline{b} \to \overline{u}u\overline{d}$
tree diagram as shown in Fig.~\ref{fig:feynman}.  Interference between the direct decay and decay
after $\BBbar$ mixing in $\Bz \to \pi^+\pi^-$ results in a time-dependent decay rate asymmetry
 that is sensitive to the angle $\alpha$.  The presence of gluonic loop (penguin)
contributions with a different weak phase to the tree contribution shifts the
measured angle from the Unitarity Triangle angle $\alpha$ to an effective parameter $\alpha^{\pi\pi}_{\mathrm eff}$,
where the shift is defined as $\delta\alpha^{\pi\pi} = \alpha-\alpha^{\pi\pi}_{\mathrm eff}$.

The time-dependent \CP\ asymmetry measured in $\Bz \to \pi^+\pi^-$ has the
form\footnote{The Belle Collaboration measure \acp\ by convention, where $\acp=-\C$.}
\begin{equation}
{\calA}=S\sin(\deltamd\deltat)-\C\cos(\deltamd\deltat),\label{eqn:asym}
\end{equation}
where $\deltamd$ is the $\BBbar$ mixing frequency, $\deltat$ is the proper time difference between
the decay of the two \B\ mesons in an event, and
\begin{eqnarray}
\S = \frac{ 2 \Im (\lambda)     }{ 1 + |\lambda|^2}, \hspace{2.0cm}
\C = \frac{ 1 - |\lambda|^2 }{ 1 + |\lambda|^2},
\end{eqnarray}
where $\lambda= (q/p)(\overline{A}/A)$.  The ratio $q/p$ is related to
\CP\ violation in mixing of neutral \B\ mesons, and $\overline{A}/A$ is the ratio
of decay amplitudes of $\Bzb\to\pi^+\pi^-$ to $\Bz\to\pi^-\pi^+$ decays.  If penguin
amplitudes contribute to the decay, $\S=\sqrt{1-\C^2}\sin(2\alpha^{\pi\pi}_{\mathrm eff})$,
and \C\ can be non-zero.

\begin{figure}[!h]
\begin{center}
  \resizebox{7.5cm}{!}{\includegraphics{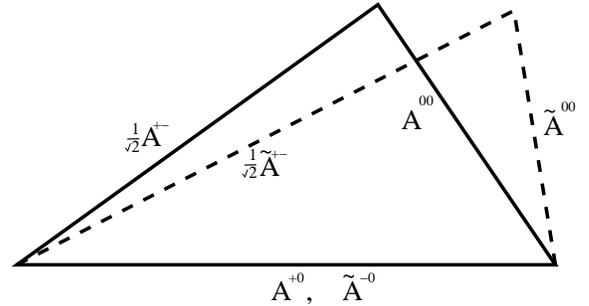}}
\end{center}
 \caption{The isospin triangle for $\B\to\pi\pi$ decays.}
\label{fig:ispintriangle}
\end{figure}

Gronau and London proposed a method using \su{2} isospin symmetry to cleanly
disentangle the penguin contribution and extract $\alpha$ from measurements
of \S\ and \C~\cite{gronaulondon}.  This method relates the isospin amplitudes
of $\Bz\to\pi^-\pi^+$, $\Bz\to\pi^0\pi^0$ and $\B^+\to\pi^+\pi^0$ processes and
their complex conjugates as two triangles in a complex plane. As shown in
Fig.~\ref{fig:ispintriangle},
\begin{eqnarray}
\frac{1}{\sqrt{2}}A^{+-}=A^{+0}-A^{00},\\
\frac{1}{\sqrt{2}}\overline{A}^{+-}=\overline{A}^{-0}-\overline{A}^{00},
\end{eqnarray}
where $A^{ij}$ ($\overline{A}^{ij}$) are the amplitudes of \B\
($\overline{B}$) decays to the final state with charge $ij$, and $\tilde{A}^{ij}$ denotes
a rotated amplitude $\overline{A}^{ij}$.
 After aligning
the amplitudes for $\pi^+\pi^0$ and $\pi^-\pi^0$ decays, the phase difference
between the $A^{+-}$ and $\tilde{A}^{+-}$ is a measure of
$2\delta\alpha^{\pi\pi}$.  The theoretical uncertainties in this method are at a level
of a few degrees and are discussed at length in Ref.~\cite{melicCKM}.

\section{Experimental Techniques}

\subsection{Signal Extraction}

Continuum $\epem \to \qqbar$ ($q = u,d,s,c$) events are the
dominant background to $B\to \pi\pi$ decays. Signal
candidates are identified using two kinematic variables: the difference
\deltae\ between the energy of the \B\ candidate and the beam
energy $\sqrt{s}/2$ in the center of mass frame; and the
beam-energy substituted mass
$\mes = \sqrt{(s/2 + {\mathbf {p}}_i\cdot {\mathbf {p}}_B)^2/E_i^2- {\mathbf {p}}_B^2}$.
The \B\ momentum ${\mathbf {p}_B}$ and four-momentum of the initial
state $(E_i, {\mathbf {p}_i})$ are defined in the laboratory frame.
Event shape variables are used to further discriminate between
signal and continuum background.  The selection of $\Bz\to\pi^+\pi^-$ decays
also depends on particle identification information to discriminate between the signal
$\pi^+\pi^-$ and background $K^\pm\pi^\mp$ decays.

\subsection{Recent improvements in analysis techniques}

The Belle analysis of $\Bz\to\pi^+\pi^-$ has recently been improved by using particle
identification information from the aerogel cherenkov counters and $\mathrm{dE/dx}$
from the drift chamber to discriminate between charged $\pi$ and $K$ mesons.
The \babar\ analysis of $\Bz\to\piz\piz$ and $\B^\pm \to \pi^\pm\piz$ has recently
benefited from an improved \piz\ meson selection.  The original selection technique used
in these analyses reconstructed $\piz\to\gamma\gamma$ decays where the final state photons
deposit all of their energy in distinct regions of the \babar\ electromagnetic calorimeter.
There is a class of high energy \piz\ mesons where the two photons deposit all of their energy
in indistinguishable regions of the calorimeter.  The lateral shower profile in the calorimeter
is used to select such \piz\ mesons~\cite{babar_btopipiCP}.  In addition to this, it is possible to reconstruct
\piz\ mesons where photons undergo $\gamma \to \epem$ conversion in detector material.  Inclusion
of both of these classes of events results in an increased selection efficiency of approximately
10\% for $\Bz\to\piz\piz$.

\section{Experimental Results}

\subsection{\boldmath $\Bz \to \pi^+\pi^-$}

\begin{figure}[!h]
\begin{center}
% -0.59 ± 0.09    -0.39 ± 0.07    -0.10
  \resizebox{4.5cm}{!}{\includegraphics{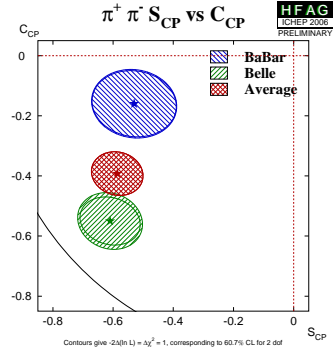}}
\end{center}
 \caption{The $S$ and $C$ measurements for $\B\to\pi^+\pi^-$ decays.}
\label{fig:scplot}
\end{figure}

\begin{figure}[!tb]
\begin{center}
  \resizebox{6.0cm}{!}{\includegraphics{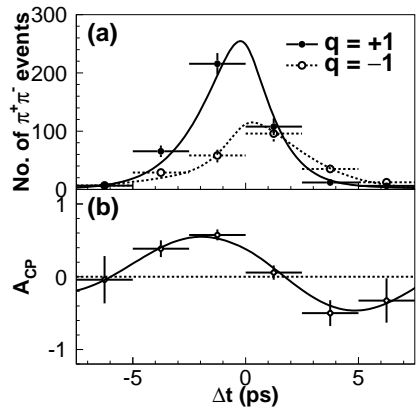}}
\end{center}
\caption{(top) The $\Delta t$ distribution for (solid) \Bz\ and (dashed)
 \Bzb\ tagged $\Bz\to \pi^+\pi^-$ decays and (bottom) the time-dependent $\CP$ asymmetry from Belle.}
\label{fig:belleasymmetry}
\end{figure}

Table~\ref{tbl:pippim} summarises the experimental measurements of $\Bz\to \pi^+\pi^-$ decays reported
in Refs.~\cite{babar_btopipiCP,babar_btopipiBF,belle_btopipiCP,belle_btopipiBF}.  The values for
$C$ measured by Belle and \babar\ differ by 2.3$\sigma$, and all other measurements are very compatible
between the two experiments as shown in Fig.~\ref{fig:scplot}.  The \babar\ (Belle) Collaborations
exclude the possibility of no \CP\ violation with both $S=0$ and $C=0$ at a level of 3.6 ($>6$) $\sigma$.  The Belle data constitute an observation
of direct, and mixing induced \CP\ violation in \B\ decays. Fig.~\ref{fig:belleasymmetry} is from Belle
(Ref.~\cite{belle_btopipiCP}) and shows the $\deltat$ distributions
of \Bz\ and \Bzb\ tagged events, and {$\cal{A}$}.

When calculating the branching fraction, both collaborations have accounted for possible final
state radiation in this decay.

\begin{table}[!tb]
\caption{Experimental constraints on $\Bz \to \pi^+\pi^-$ decays.
  The last column indicates the number of \BBbar\ pairs analysed.}\label{tbl:pippim}
\begin{center}
 \begin{tabular}{l|lc}\hline
Experiment & Observable & $N_{\BBbar} \e{6}$ \\ \hline
Belle      & ${\cal B} = (5.1 \pm 0.2 \pm 0.2)\e{-6}$ & 449\\
           & $C = 0.55 \pm 0.08 \pm 0.05$             & 535\\
           & $S = -0.61 \pm 0.10 \pm 0.04$            & 535\\ \hline
\babar     & ${\cal B} = (5.8 \pm 0.4 \pm 0.3)\e{-6}$ & 227\\
           & $C = -0.16 \pm 0.11 \pm 0.03$            & 347\\
           & $S = -0.53 \pm 0.14 \pm 0.02$            & 347\\ \hline
 \end{tabular}
\end{center}
\end{table}

\subsection{\boldmath $\B^+ \to \pi^+\pi^0$}

Table~\ref{tbl:pippiz} summarises the experimental measurements of $\B^+\to \pi^+\pi^0$ decays reported
in Refs.~\cite{babar_btopipiCP,belle_btopipiBF}.  In the absence of electroweak penguins (EWP)
 $\acp=0$ in this channel.  Experimental constraints are consistent with these expectations.
Estimates of the possible theoretical uncertainty from EWP contributions to $\B^+ \to \pi^+\pi^0$
on $\alpha$ are at the level of a few degrees~\cite{zupanCKM,melicCKM,ckmfitter}.

\begin{table}[!tb]
\caption{Experimental constraints on $\B^+ \to \pi^+\pi^0$ decays.
  The last column indicates the number of \BBbar\ pairs analysed.}\label{tbl:pippiz}
\begin{center}
 \begin{tabular}{l|lc}\hline
Experiment & Observable & $N_{\BBbar} \e{6}$                 \\ \hline
Belle      & ${\cal B} = (6.6 \pm 0.4 ^{+0.4}_{-0.5})\e{-6}$ & 535\\
           & $\acp = 0.07 \pm 0.06 \pm 0.01$                 & 535\\ \hline
\babar     & ${\cal B} = (5.12 \pm 0.47 \pm 0.29)\e{-6}$     & 347\\
           & $\acp = -0.019 \pm 0.088 \pm 0.014$             & 347\\
 \end{tabular}
\end{center}
\end{table}

\subsection{\boldmath $\Bz \to \pi^0\pi^0$}

Table~\ref{tbl:pizpiz} summarises the experimental measurements of $\Bz\to \pi^0\pi^0$ decays reported
in Refs.~\cite{babar_btopipiCP,belle_bztopizpiz}.
The measurement of $C$ in this mode is an important
contribution to the isospin analysis as the measured $\B\to \pi\pi$ branching fractions, with
$C$ from $\pi^+\pi^-$ and $\piz\piz$ give as many constraints as observables required to fully
constrain the isospin analysis.
Experimental knowledge of this decay is the limiting factor in improving the constraint on $\alpha$
in $\B\to \pi\pi$ decays using an isospin analysis.

\begin{table}[!tb]
\caption{Experimental constraints on $\Bz \to \pi^0\pi^0$ decays.
  The last column indicates the number of \BBbar\ pairs analysed.}\label{tbl:pizpiz}
\begin{center}
 \begin{tabular}{l|lc}\hline
Experiment & Observable & $N_{\BBbar} \e{6}$ \\ \hline
Belle      & ${\cal B} = (1.1 \pm 0.4 \pm 0.1)\e{-6}$        & 535\\
           & $C = -0.44 ^{+0.62}_{-0.73} \,^{+0.06}_{-0.04}$ & 535\\ \hline
\babar     & ${\cal B} = (1.48 \pm 0.26 \pm 0.12)\e{-6}$     & 347\\
           & $C = -0.33 \pm 0.36 \pm 0.08$                   & 347\\
 \end{tabular}
\end{center}
\end{table}

\subsection{Results of the Isospin Analysis}

\begin{figure}[!tb]
\begin{center}
  \resizebox{7.cm}{!}{\includegraphics{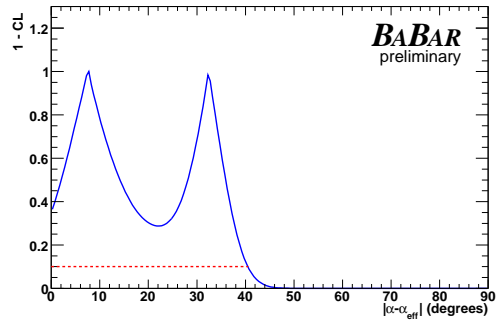}}
\end{center}
 \caption{The current $\delta\alpha^{\pi\pi}$ constraint from \babar.}
\label{fig:IAprediction:deltaalpha}
\end{figure}

\begin{figure}[!tb]
\begin{center}
  \resizebox{7.cm}{!}{\includegraphics{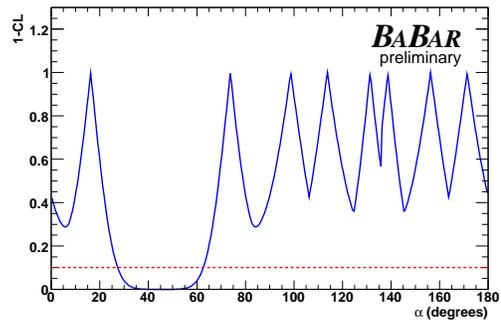}}
  \resizebox{7.cm}{!}{\includegraphics{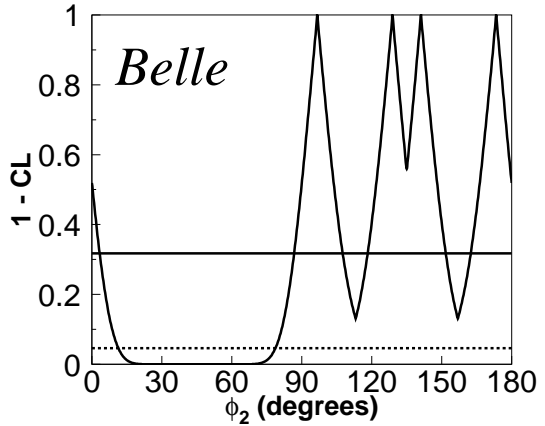}}
\end{center}
 \caption{The current \su{2} analysis 1-CL constraint on $\alpha$ from (top) \babar (bottom) Belle.}
\label{fig:IAresult}
\end{figure}

Fig.~\ref{fig:IAprediction:deltaalpha} shows the one minus confidence level (1-CL) distribution for
$|\delta \alpha^{\pi\pi}|$ obtained from
an isospin analysis of $\B \to \pi\pi$ decays.  The upper limit from penguin pollution in the measurement of
$\alpha$ from $\B \to \pi\pi$ is $41^\circ$ (90\%) CL from \babar. The constraint from Belle is tighter as
the \piz\piz\ branching fraction from Belle is smaller than the one from \babar. There is
a dip on the 1-CL distribution between
$|\delta \alpha|\sim 10^\circ$ and $|\delta \alpha|\sim 35^\circ$ (See Fig.~\ref{fig:IAprediction:deltaalpha})
which is the result of including the measurement
of $C$ from $\Bz \to \piz\piz$ decays in the isospin analysis.  The 1-CL distribution for $\alpha$ obtained
from an isospin analysis of $\B \to \pi\pi$ decays is shown in Fig.~\ref{fig:IAresult}.  Values of $\alpha$
between $11^\circ$ and $79^\circ$ are excluded at 95.4\% CL by Belle, and weaker constraints are obtained by \babar.
The main reason for the difference between the constraints reported by the \B-factories is that the \babar\ paper
uses only \babar\ results, whereas the Belle paper uses the average of the \babar\ and Belle results.  In the latter
case the isospin triangles are flatter than in the former case.  The effect this has on the 8-fold ambiguity on
$\alpha$ from an isospin analysis is to overlap adjacent solutions.

\section{Sensitivity Projections}

\begin{figure}[!tb]
\begin{center}
% See the release AWGclsq2b01/bevan/hh/analysis-32
  \resizebox{7.cm}{!}{\includegraphics{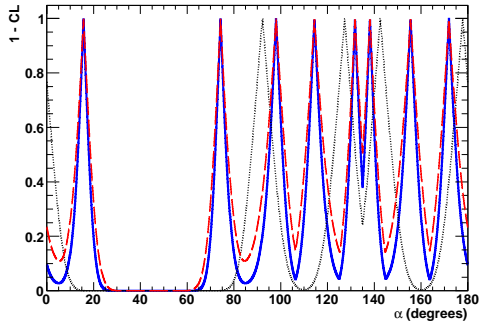}}
\end{center}
 \caption{Predictions of the \su{2} analysis 1-CL distribution as a function of $\alpha$ obtainable from the \B-factories for
          data samples of (solid) 2\iab and (dashed) 1\iab.  These extrapolations use the \babar\ central values as
          inputs to the isospin analysis. The dotted curve corresponds to the prediction
          obtained when extrapolating the current world average values of $B\to \pi\pi$ measurements to 2\iab.}
\label{fig:IAprediction}
\end{figure}

Using the current experimental knowledge of $\B\to\pi\pi$ decays, it is possible to make predictions of the
constraint on $\alpha$ from a \B-factory with 1\iab\ of data, and by combining the results of Belle and \babar\
for a total data sample of 2\iab.  These predictions use the central values given by \babar, and scale the
statistical and the subset of reducible systematic uncertainties to higher luminosity, and have been made using the
statistical technique described in~\cite{babarrhoprhomr14}.
The results of the isospin analysis
predictions for 1\iab\ and 2\iab\ are shown in Fig.~\ref{fig:IAprediction}.
The precision
of a single SM solution from 1\iab\ (2\iab) of data is expected to be $\pm 6^\circ$ ($\pm 4.2^\circ$) at 68\% CL, which remains
larger than current theoretical uncertainties from the isospin analysis.  If one
performs an isospin analysis projection using the world average of the inputs (also shown), then the
precision of each distinct solution is degraded by a few degrees.  As a result the eight-fold
ambiguity in the solution for alpha is manifest as four distinct solutions for $\alpha$ between
zero and $180^\circ$.

It is not possible to select between the different ambiguities of the isospin analysis with $\B\to\pi\pi$ decays,
and one will ultimately have to combine the constraints obtained from $\pi\pi$,
$\rho\pi$~\cite{babar_btorhopi,belle_btorhopi} and $\rho\rho$~\cite{babarrhoprhomr15prelim,babarrhoprhoz,babarrhozrhoz,bellerhoprhom,bellerhorho0}
for the final \B-factory measurement of $\alpha$.

 As the $\piz\piz$ final state is difficult to isolate in a hadronic environment, it is expected that a
 Super Flavor Factory operating at an $\epem$ collider will be required to significantly improve
 upon the \B-factory isospin analysis constraint on $\alpha$ from $\pi\pi$ decays.
 It might become feasible to perform a time-dependent analysis of $\Bz\to\piz\piz$ at a future Super Flavor Factory.
 This would require reconstruction of the decay vertex of the $\piz\piz$ final state where the \piz\ mesons
 undergo conversion in detector material, or Dalitz decay.
 If this was achievable, the measurement of $S$ in $\Bz \to \piz\piz$ could be used to overconstrain
 the $\pi\pi$ isospin analysis.

\section{Summary}

The \B-factories have seen compelling evidence for \CP\ violation in $\B \to \pi^+\pi^-$ decays. The Belle data are
compatible with observations of both direct and mixing-induced \CP\ violation signals. The \babar\ data
provides weaker evidence for such effects.  There is still a small discrepancy between the two experiments
in the measurement of $C$ in this channel, however it is expected that
the two experiments will continue to move toward a common value with more data.  The $\alpha$ constraint
obtained from an isospin analysis of $\pi\pi$ decays at 68\% CL is becoming increasingly precise.
A single solution could be measured to $4.2^\circ$ at 68\% CL, with a data sample of 2\iab, at the end
of the lifetime of the \B-factories.  This would provide a non-trivial constraint on $\alpha$ after
combination with the available information from $\rho\rho$ and $\rho\pi$ decays.

\bibliographystyle{unsrt}
\bibliography{biblio}
\end{document}